%% file: main.tex
\documentclass[conference]{IEEEtran}
\IEEEoverridecommandlockouts
\usepackage{cite}
\usepackage{amsmath,amssymb,amsfonts}
\usepackage{algorithmic}
\usepackage{graphicx}
\usepackage{textcomp}
\usepackage{multirow}
\usepackage{subcaption}
\usepackage{url}
\usepackage{graphicx}
\usepackage[table,xcdraw]{xcolor}

\def\BibTeX{{\rm B\kern-.05em{\sc i\kern-.025em b}\kern-.08em
    T\kern-.1667em\lower.7ex\hbox{E}\kern-.125emX}}
\begin{document}

\newcommand{\yn}[1]{{\color{green}{[Youness: #1]}}}
\newcommand{\mb}[1]{{\color{orange}{[Matteo: #1]}}}
\newcommand{\lc}[1]{{\color{red}{[Luca: #1]}}}
\newcommand{\id}[1]{{\color{purple}{[Idilio: #1]}}}
\newcommand{\mm}[1]{{\color{blue}{[Marco: #1]}}}

\title{Autonomous LLM Agents \& CTFs: A Second Look}

\author{
Youness Bouchari, Matteo Boffa, 
Marco Mellia\\
\textit{Politecnico di Torino}\\
\{first.last\}@polito.it
\and
Idilio Drago\\
\textit{Università di Torino}\\
idilio.drago@unito.it
\and
Thanh Minh Bui, Dario Rossi\\
\textit{Huawei Technologies France}\\
\{first.last\}@huawei.com
}

\maketitle
\begin{abstract}
\input{sections/00_abstract}
\end{abstract}
\section{Introduction}
\input{sections/01_introduction}
\section{LLM-based Agents: Syllabus}
\input{sections/02_agent_syllabus}
\section{Methodology and Dataset}
\input{sections/03_methodology}
\section{Results}\label{sec:results}
\input{sections/04_results}

\section{Conclusion and Future Work}

\input{sections/05_conclusion}
\bibliographystyle{IEEEtran}
\bibliography{main}
\newpage
\section*{Appendix}
\input{sections/06_appendix}

\end{document}

%% file: sections/00_abstract.tex
Large Language Model (LLM) agents are increasingly proposed to automate offensive security tasks, with recent studies reporting near human-level success rates in Capture-the-Flag (CTF) challenges. We here revisit these results, providing a \emph{second look} at these claims. We engineer different agent architectures of increasing complexity and modularity on 30 web-based CTFs challenges spanning 14 vulnerability classes. We instantiate these agents with multiple LLM backbones, and compare them with \texttt{claude-code}, a general-purpose agent that automatically determines its internal architecture.
Our evaluation yields three main findings. First, \texttt{claude-code} achieves performance comparable to the engineered architectures (19/30 solved tasks), suggesting that general-purpose agents are strong baselines for offensive security tasks. Second, both our architectures and \texttt{claude-code} struggle in the same challenge categories, revealing persistent barriers that keep current agents below human-level capability. Third, by leveraging our manually designed architectures we can systematically measure the impact of additional components, finding that structured orchestration of specialized roles outperforms monolithic designs, improving run-to-run consistency, and reducing execution costs.

%% file: sections/01_introduction.tex
\emph{Penetration testing} is a systematic security evaluation to identify and exploit vulnerabilities in a target system~\cite{bishop2007, nist_sp800115}. Despite its importance for proactive cyber defense, it remains difficult to automate. Nowadays, the global cybersecurity workforce lacks nearly 4.8 million professionals~\cite{isc22024workforce}, while attackers often start exploiting newly disclosed vulnerabilities within 15 minutes from the CVE announcements~\cite{unit42_2025_global_incident_response}. 

\emph{Large Language Models} (LLMs) are promising candidates for automating this process, given their knowledge of vulnerability classes, attack techniques, and security tools~\cite{zhang2025llms}. However, offensive security requires multi-step planning, adaptive tool use, and coherent context management over long interaction horizons~\cite{deng2024pentestgpt} -- capabilities that current LLM-based agents are only starting to demonstrate. 
Early work showed that LLMs can assist penetration testers in tool invocation, output interpretation, and action planning~\cite{happe2023, deng2024pentestgpt, gioacchini2025autopenbench}. Multi-agent architectures -- where a central planner coordinates specialized sub-agents -- have since outperformed monolithic approaches across network and web application settings~\cite{zhu2024zeroday, kong2025vulnbot, david2025mapta}. Dedicated benchmarks now standardize evaluation, spanning CTF-style suites~\cite{zhang2025cybench, gioacchini2025autopenbench} and sandboxed real-world CVEs~\cite{zhu2025cvebench}, with agentic systems reporting near-human success rates~\cite{david2025mapta, anthropic2025cybercompetitions, xbow2025top1, lin2026comparing}. Nevertheless, many approaches remain closed-source or assess proxy tasks (e.g., vulnerability recognition~\cite{david2025mapta}) rather than full exploit execution.

We conduct a \emph{reality check} on LLM-based multi-agent systems for offensive security, providing a \emph{second-look} at their current potential in pentest-like tasks. Can prior claims be confirmed? Where do LLM agents still struggle? Do hand-crafted multi-agent architectures offer tangible advantages over general-purpose agentic tools? We focus on \emph{web application} penetration testing -- accounting for nearly 44\% of security incidents~\cite{unit42_2025_global_incident_response} -- and use web-based Capture the Flag (CTF) challenges as safe and reproducible proxies for real attack scenarios~\cite{vigna2014ten,shao2024nyu}.

\begin{figure*}[t]
\centering

\begin{subfigure}[t]{0.31\textwidth}
    \centering
    \includegraphics[width=\linewidth]{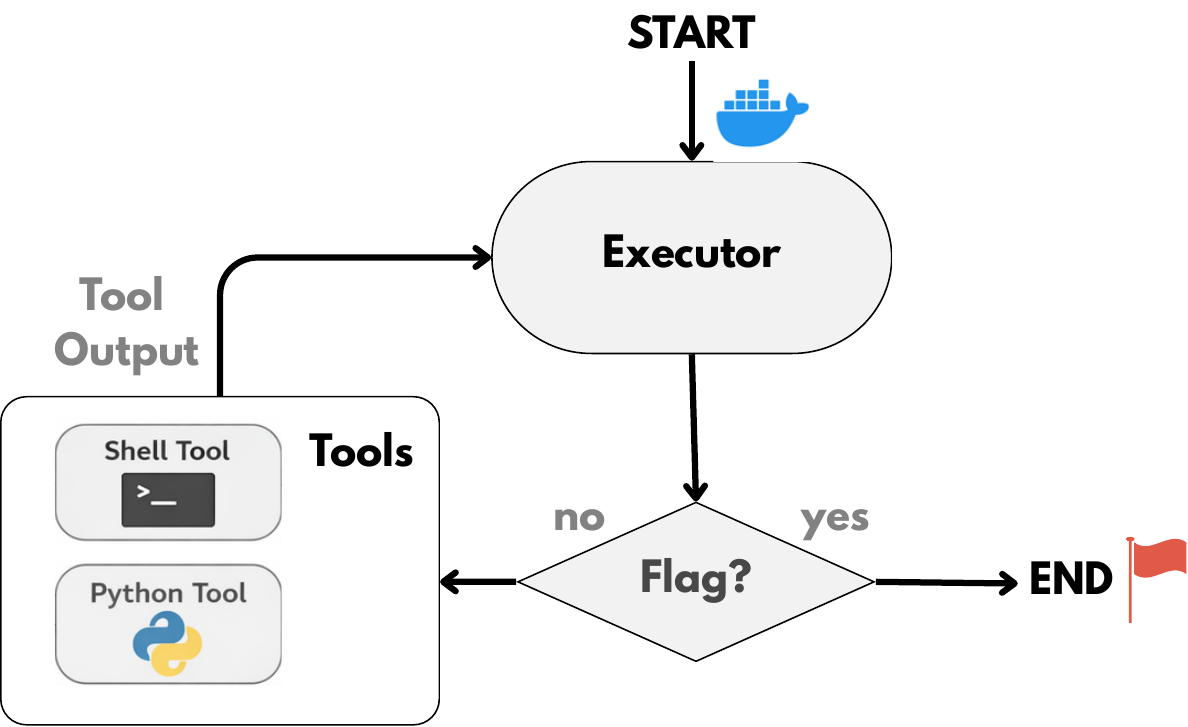}
    \caption{Executor}
    \label{fig:arch_executor}
\end{subfigure}
\hfill
\vrule width 0.5pt
\hfill
\begin{subfigure}[t]{0.31\textwidth}
    \centering
    \includegraphics[width=\linewidth]{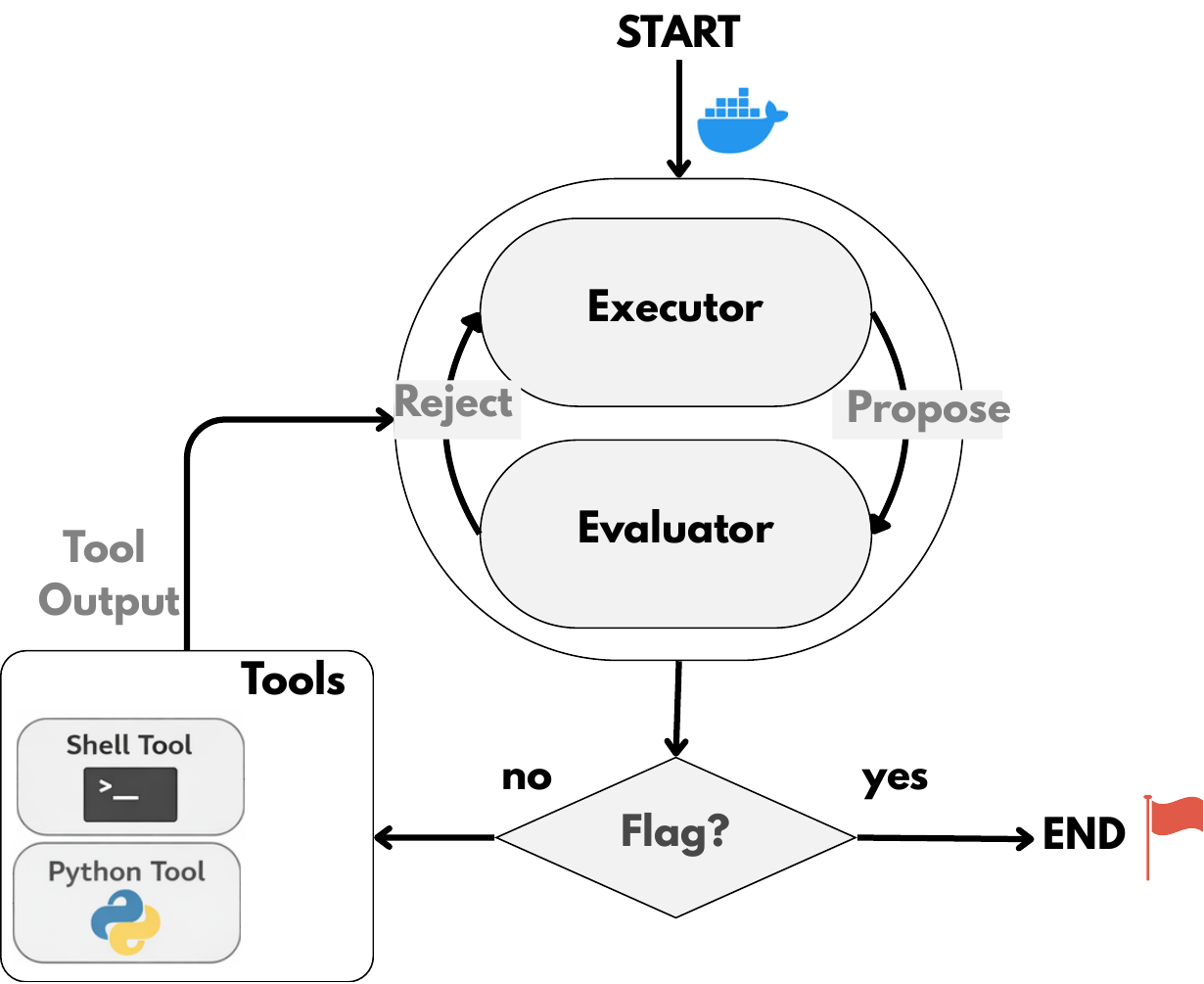}
    \caption{Executor + Evaluator}
    \label{fig:arch_eval}
\end{subfigure}
\hfill
\vrule width 0.5pt
\hfill
\begin{subfigure}[t]{0.31\textwidth}
    \centering
    \includegraphics[width=\linewidth]{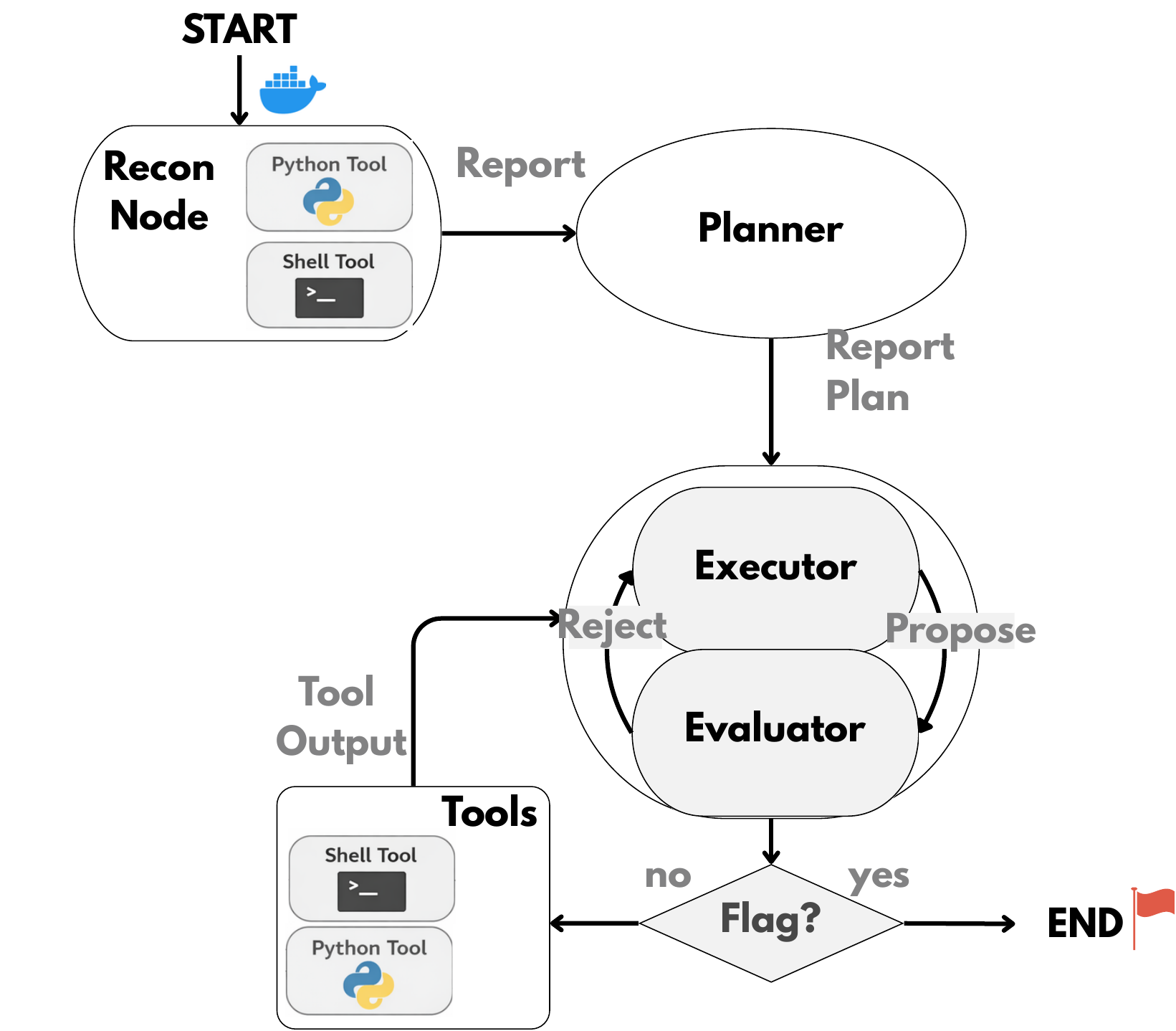}
    \caption{Planner + Executor + Evaluator}
    \label{fig:arch_full}
\end{subfigure}

\caption{\textbf{Tested agent architectures.} We progress from a single-agent \emph{Executor}~(a) to a structured multi-agent configuration~(c) consisting of a \emph{Recon Node}, \emph{Planner}, and \emph{Evaluator}. We use \texttt{claude code} (not shown) as a baseline. All systems are granted access to a vulnerable (Dockerized) service and terminate by outputting the flag, if successfully captured.}
\label{fig:architectures}
\end{figure*}

We design and evaluate LLM agents of increasing complexity on a curated set of 30 CTF challenges 
covering 14 vulnerability classes, with no public solutions available. We manually solve all challenges to enable fine-grained analysis of agent reasoning and failure patterns. We introduce three domain-specific architectures (Figure~\ref{fig:architectures}) that combine an \emph{Executor} (environment interaction), an \emph{Evaluator} (iterative plan refinement) and a \emph{Planner} (vulnerability classification and attack strategy). We also benchmark against \texttt{claude-code}, a production-grade agentic assistant that can autonomously spawn sub-agents, using our transparent pipelines as interpretable counterparts for failure analysis.

Our evaluation yields three main findings.
First, both our best architecture and \texttt{claude-code} solve 19/30 challenges (63\%), with our engineered solution being more efficient (fewer steps and lower cost), revealing that current agents are still below human-level capability. 
Second, we find that failures consistently cluster around the same set of tasks -- including business logic flaws, race conditions, and blind SQL injection -- and we identify the technical and semantic barriers underlying these persistent failures.
Finally, we use our manually designed architectures to study how architectural components affect agent behaviour. We find that single monolithic agents frequently produce locally plausible actions but lack a global plan, whereas multi-agent systems -- particularly those including a Planner -- achieve solutions more systematically.

To facilitate reproducibility and future research, we publicly release our agent implementations and challenge traces.\footnote{\url{https://github.com/SmartData-Polito/CTF_agent}}

%% file: sections/02_agent_syllabus.tex
We introduce the core concepts of LLM-based agents and refer the reader to recent surveys for more comprehensive overviews~\cite{wang-etal-2023-plan, AgentSurveyKDD25}.

An \emph{agent} is an autonomous system that pursues goals by interacting with an \emph{environment} through \emph{actions} that alter the \emph{state} of the environment~\cite{sumers2023cognitive}. An LLM can function as the decision-making engine, supporting iterative reasoning and action selection rather than one-shot generation~\cite{yao2023react}.

In the context of penetration testing, the environment consists of the target infrastructure (e.g., a web application) and the associated tooling ecosystem (e.g., an exploitation framework). At each step, the agent i) executes an \emph{action} (e.g., crafting an exploit), ii) receives an \emph{observation} (e.g., command outputs), iii) updates its internal state, and iv) selects the next action according to its \emph{policy}. For most LLM-based agents, this policy is implemented with \emph{prompts}. A prompt is the structured textual input provided to the LLM. The prompt encodes the current context (e.g., prior observations, goals, constraints) and guides the next action or decision from the model~\cite{yao2023react}.

More formally, following the CoALA cognitive architecture for language agents~\cite{sumers2023cognitive}, an LLM-based agent can be decomposed into the following core components:

\begin{itemize}
\item \textbf{Reasoning and Decision-Making:} The LLM iteratively interleaves the reasoning and acting steps, forming a thought--action--observation loop that supports long-horizon problem solving~\cite{yao2023react}.

\item \textbf{Actions and Tool Use:} The agent interacts with the environment through predefined tools (e.g., APIs, command execution). Tool-augmented language models improve reliability by delegating well-defined sub-tasks to external systems~\cite{schick2023toolformer}.

\item \textbf{Memory:} The agent maintains an internal state through the steps. The LLM's memory is embedded into a \emph{scratchpad} stored within the prompt, where intermediate reasoning traces and observations are accumulated~\cite{yao2023react}. This corresponds to short-term working memory in CoALA~\cite{sumers2023cognitive}.
\end{itemize}

The agent is initialized through a prompt that specifies the task, available tools, and operational constraints. As the interaction progresses, the scratch-pad grows with reasoning traces, executed actions, and observations. This design is bounded by the LLM's \emph{context window}, which limits the amount of information the LLM can process simultaneously.

LLM agents can be extended to \emph{multi-agent} architectures, where specialized sub-agents (e.g., \textit{recon node}, \textit{executor}, \textit{evaluator}) collaborate under a coordinating agent~\cite{wu2023autogen, chen2023agentverse}. While modularisation improves scalability, orchestration becomes critical to ensure coherence and avoid error propagation. Hence, structured orchestration can improve stability and performance. For example, separating planning from execution -- i.e., enforcing a pipeline where a \emph{planner} first produces a high-level attack plan and an \emph{executor} follows it without iterative re-planning -- constrains the search space of the acting policy and reduces oscillatory behaviours~\cite{wang-etal-2023-plan}. 

%% file: sections/03_methodology.tex
\subsection{Agent Design}

We introduce the agent architectures in Figure~\ref{fig:architectures}, starting with common traits (e.g., prompt, tools, and memory) and then moving on to their differences.

\textbf{Common Traits}: All agents share the following:

$\bullet$ \emph{Prompt:} We use a common prompt to define the role of the agent (`Capture the Flag (CTF) expert') and task (`identify the vulnerability and capture the flag'). We also provide instructions on available tools (next point), general guidance (`be systematic and grounded'), operational constraints (`do not browse the host filesystem'), and exit conditions: output `\texttt{FLAG \{...\}}' if the flag is found; otherwise `\texttt{GIVE\_UP}' if you believe the challenge is unsolvable.

$\bullet$ \emph{Tools:} Agents interact with the environment (the vulnerable endpoints) through an SSH terminal. Specifically, we provide two tools: \texttt{run\_command}, i.e., directly run any bash command, and \texttt{run\_python}, allowing the agent to create and run python scripts. We instruct the agents to prefer \texttt{run\_command} for simple explorations and inspections and to save \texttt{run\_python} for structured programs. In both cases, we ask the model to couple the tool call with a \texttt{reason} explaining the inner thought process that led to the call.

$\bullet$ \emph{Memory Management:} The agents track previous steps by appending reasoning traces, tool calls, and corresponding output into the scratchpad. As LLMs only have a limited context, we ask the agent to summarize previous findings to produce more concise reports if the latter becomes too long. 

\textbf{Domain-Specific Architectures}: We design three domain-specific agent systems on top of the previous building blocks:

$\bullet$ \emph{Executor (E):} The simplest configuration in which a single monolithic agent is responsible for i) discovering the vulnerable environment, ii) drafting an exploitation plan, and iii) executing it. Previous work~\cite{fumero2025cybersleuth} showed that this design tends to overburden the agent, leading to degraded performance.

$\bullet$ \emph{Executor + Evaluator (E+E):} This multi-agent architecture augments the Executor with an Evaluator to iteratively refine its actions. The Evaluator operates in an LLM-as-a-judge setting~\cite{li2025generation}, scoring the Executor's \texttt{reasons} before tool calls are executed. If the score is low, the Evaluator blocks the execution and suggests the Executor to retry, providing text-based feedback for improvement.
Although in-the-loop evaluation improves step-level efficiency, the Evaluator often makes judgments without sufficient context: it assesses each action in isolation, without visibility into the full execution trace, and therefore cannot reliably determine whether the overall execution logic is coherent and consistent.

$\bullet$ \emph{Planner + Executor + Evaluator (P+E+E):} This architecture introduces a Planner, supported by a reconnaissance node, which first explores the environment and formulates a plan. The plan is then provided to both the Executor and the Evaluator. This design (i) decouples planning from execution, simplifying the Executor’s task, and (ii) equips the Evaluator with a structured reference, enabling more informed decisions when accepting or rejecting actions.

\textbf{General-Purpose Agent}: As an external baseline, we evaluate \texttt{claude-code}, an autonomous agentic system that has recently been reported to perform competitively with humans in cybersecurity competitions~\cite{anthropic2025cybercompetitions}.
Unlike our explicitly modular architectures (E, E+E, P+E+E), \texttt{claude-code} is a self-directed coding agent that autonomously determines how to approach a task. It interacts with the environment through a set of primitive tools -- such as \texttt{bash} (shell execution), \texttt{read}, \texttt{write}, and \texttt{edit} -- which it uses to inspect files, generate and refine code, and execute multi-step exploitation strategies.
Importantly, \texttt{claude-code} supports autonomous multi-agent instantiation. Specifically, the main agent can autonomously spawn \emph{subagents} -- specialized instances operating in independent context windows with custom system prompts and restricted tool access -- and delegate focused subtasks to them. In addition, it uses a native tool-use framework and internal memory mechanisms~\cite{claudecode_docs_overview, claudecode_docs_memory}.
Overall, \texttt{claude-code} serves as a strong, general-purpose agent baseline against which we compare our explicitly structured modular designs.

\subsection{Benchmark}

We evaluate our methods against a subset of challenges from the XBOW benchmark~\cite{xbow2025top1}. XBOW is a curated collection of web-based CTF tasks designed to assess automated vulnerability detection and exploitation systems. The benchmark covers a diverse range of realistic web security scenarios, spanning multiple vulnerability categories and difficulty levels.

The benchmark authors do not release official solutions to reduce the risk of data leakage in newer LLMs. We manually solve a subset of \emph{30 challenges} that we use as reference solutions to enable interpretability and fine-grained evaluation of whether and at what stage agents perform the correct exploitation steps.\footnote{Following the benchmark authors’ policy, we do not release these solutions in our repository and make them available upon request.}

Table~\ref{tab:xbow_distribution} summarises the characteristics of the 30 selected challenges in vulnerability categories and difficulty levels. We intentionally sample tasks to cover a broad range of vulnerability types and complexities. Each challenge consists of an isolated Docker image that instantiate a vulnerable web application. At evaluation time, we deploy each image as a web server and provide the agent only with the corresponding URL as its entry point.

\begin{table}[t]
\centering
\caption{\textbf{Distribution of selected XBOW challenges}. We group vulnerability categories and difficulty levels.}
\label{tab:xbow_distribution}
\begin{tabular}{lccc|c}
\hline
\textbf{Vulnerability} & \textbf{Easy} & \textbf{Medium} & \textbf{Hard} & \textbf{Total} \\
\hline
Broken Cryptography              & 0 & 1 & 1 & 2 \\
Insecure Direct Object Reference & 2 & 2 & 0 & 4 \\
Insecure Design                  & 0 & 2 & 1 & 3 \\
JSON Web Token Vulnerability     & 0 & 1 & 0 & 1 \\
No SQL Injection                 & 0 & 1 & 0 & 1 \\
Path Traversal                   & 1 & 3 & 0 & 4 \\
Secure Shell-related             & 1 & 0 & 0 & 1 \\
Server-Side Request Forgery      & 1 & 0 & 0 & 1 \\
XML External Entity Injection    & 2 & 0 & 0 & 2 \\
Cross Site Scripting             & 0 & 2 & 1 & 3 \\
Command Injection                & 1 & 2 & 0 & 3 \\
Blind SQL Injection              & 1 & 0 & 0 & 1 \\
Business Logic                   & 1 & 2 & 0 & 3 \\
Race Condition                   & 0 & 0 & 1 & 1 \\
\hline
\textbf{Total}                   & 10 & 16 & 4 & 30 \\
\hline
\end{tabular}
\end{table}

%% file: sections/04_results.tex
\begin{table*}[t]
\caption{\textbf{Benchmark performance across model and agent configurations.} 
Our agents are executed 90 times in total (three runs per task), while \texttt{claude-code} is executed once per task (30 runs). 
Success (max) reports the number of tasks solved at least once across runs. Steps, Cost, and Duration are averaged over all tasks.}
\label{tab:benchmark_results}
\centering
\resizebox{\linewidth}{!}{%
\begin{tabular}{l l c c c c}
\hline
\textbf{Model} & \textbf{Setup} & \textbf{Success (max) $\uparrow$} & \textbf{Steps (avg) $\downarrow$} & \textbf{Cost (avg) $\downarrow$} & \textbf{Duration (avg s) $\downarrow$} \\
\hline
GPT-4.1 & Executor & 9/30 & 36.74 & 1.43 & \textbf{162.13} \\
\hline
\multirow{3}{*}{GPT-5} 
& Executor                        & \textbf{19/30} & 31.56 & 0.90 & 336.74 \\
& Executor + Evaluator            & \textbf{19/30} & 28.76 & 0.64 & 799.30 \\
& Planner + Executor + Evaluator  & \textbf{19/30} & \textbf{24.09} & \textbf{0.59} & 925.80 \\
\hline
Opus-4-5 & \texttt{claude-code} & \textbf{19/30} & 45.52 & 1.26 & 284.72 \\
\hline
\end{tabular}%
}
\end{table*}

\subsection{Experimental Setup and Metrics}

To evaluate the benefits of using the most recent LLMs versus previous generations, we first instantiate our architectures using GPT-4.1~\cite{openai_gpt41_2025} and GPT-5~\cite{openai_gpt5_system_card_2025} through the OpenAI API. For the baseline, we directly use the best Claude model -- Opus 4.5~\cite{anthropic_claude_opus46_2026} -- through a local installation of \texttt{claude-code}.
To account for stochasticity, we execute each architecture three times per trace, resulting in 90 runs per architecture. Due to token consumption constraints, we do not perform multiple runs for \texttt{claude-code}.\footnote{In practice, we were able to execute approximately three traces per day before reaching the usage limit.}
All runs are independent and do not share memory between executions. Each run is capped at a maximum of 50 iterations. Additionally, within a given step, the evaluator may interrupt the executor at most three times per action before execution proceeds.

To assess efficiency, reliability, and overall task performance, we report the following metrics:

\begin{itemize}
\item \textbf{Success / Failure:} A binary indicator of whether the agent successfully solves the benchmark.
\item \textbf{Steps:} The number of interaction steps required to complete the task. Here, we do not count the Evaluator calls.
\item \textbf{Cost:} The average cost per challenge in USD.
\item \textbf{Duration:} Wall-clock execution time for each run.
\end{itemize}

When multiple runs are available, we additionally measure consistency, i.e., how reliably an agent solves the same challenge across independent executions.

\subsection{Aggregate Benchmark Analysis}

Table~\ref{tab:benchmark_results} summarises the results across LLMs and agent configurations. Several observations emerge.

First, the LLM plays a decisive role. Using GPT-4.1, the agent solves only 9/30 tasks, while GPT-5 reaches 19/30. This substantial gap indicates that complex multi-step CTF challenges require stronger reasoning and planning capabilities that only last-generation LLMs provide. 

Second, all stronger systems -- GPT-5 for all configurations and \texttt{claude-code} -- succeed in 19 CTFs out of 30. This falls well short of the results reported in prior work~\cite{xbow2025top1, anthropic2025cybercompetitions}, which lacks reproducible artifacts to verify its claims. In Section~\ref{sec:analysis_errors}, we provide an in-depth analysis of agent errors and offer a possible explanation for this discrepancy. Notice that for models that we run multiple times, we report the maximum Success across runs (i.e., if the model solves the task once, we report success). 

All agents appear to be equivalent in terms of overall solvability. However, their differences lie not in \emph{whether} they solve tasks, but in \emph{how efficiently} and \emph{consistently} they do so. In fact, efficiency metrics reveal a clear architectural trend. Within the GPT-5 variants, increasing structural organization progressively reduces the average number of interaction steps (31.56 → 28.76 → 24.09). This reduction directly results in a lower average cost, making the Planner–Executor–Evaluator configuration the most cost-efficient design. In contrast, the Wall-clock duration follows a different pattern. Although the fully structured architecture improves step and cost efficiency, it introduces runtime overhead, as the planner and evaluator are instantiated at every execution regardless of task complexity -- a cost simpler architectures avoid. \texttt{claude-code}, by contrast, rarely spawns additional agents and relies almost exclusively on \texttt{read} and \texttt{bash} tools, making its runtime roughly comparable to the bare \textit{Executor}, further aided by product-level optimizations.

\subsection{Comparing the Proposed Architectures}

\textbf{Consistency Analysis:} As previously observed, more sophisticated architectures improve overall consistency across runs. In Table~\ref{tab:architecture_consistency}, we summarize the number of challenges consistently solved by each architecture in three independent runs. All architectures use GPT-5 as the backend model.

\begin{table}[h]
\caption{\textbf{Success consistency of the proposed architectures}. We run each architecture for three independent runs.}
\label{tab:architecture_consistency}
\centering
\resizebox{\columnwidth}{!}{
\begin{tabular}{l c c c | c}
\hline
\textbf{Architecture} & \textbf{1/3} & \textbf{2/3} & \textbf{3/3} & \textbf{Total / 90} \\
\hline
Executor only                  & 5 & 2 & 12 & 45 \\
Executor + Evaluator           & 2 & 3 & 14 & 50 \\
Planner + Executor + Evaluator & 1 & 2 & 16 & \textbf{53} \\
\hline
\end{tabular}
}
\end{table}

The results show a clear trend: while all architectures achieve similar \emph{best-case} performance (19/30), robustness improves with structural complexity. The number of tasks solved in all three runs increases (12 $\rightarrow$ 14 $\rightarrow$ 16), as does the total number of successful runs (45 $\rightarrow$ 50 $\rightarrow$ 53). 
This confirms that multi-agent architectures enhance reliability rather than peak capability. The Planner--Executor--Evaluator setup reduces variance and yields the most stable overall performance.

\begin{figure}[hb]
\centering
\includegraphics[width=\columnwidth]{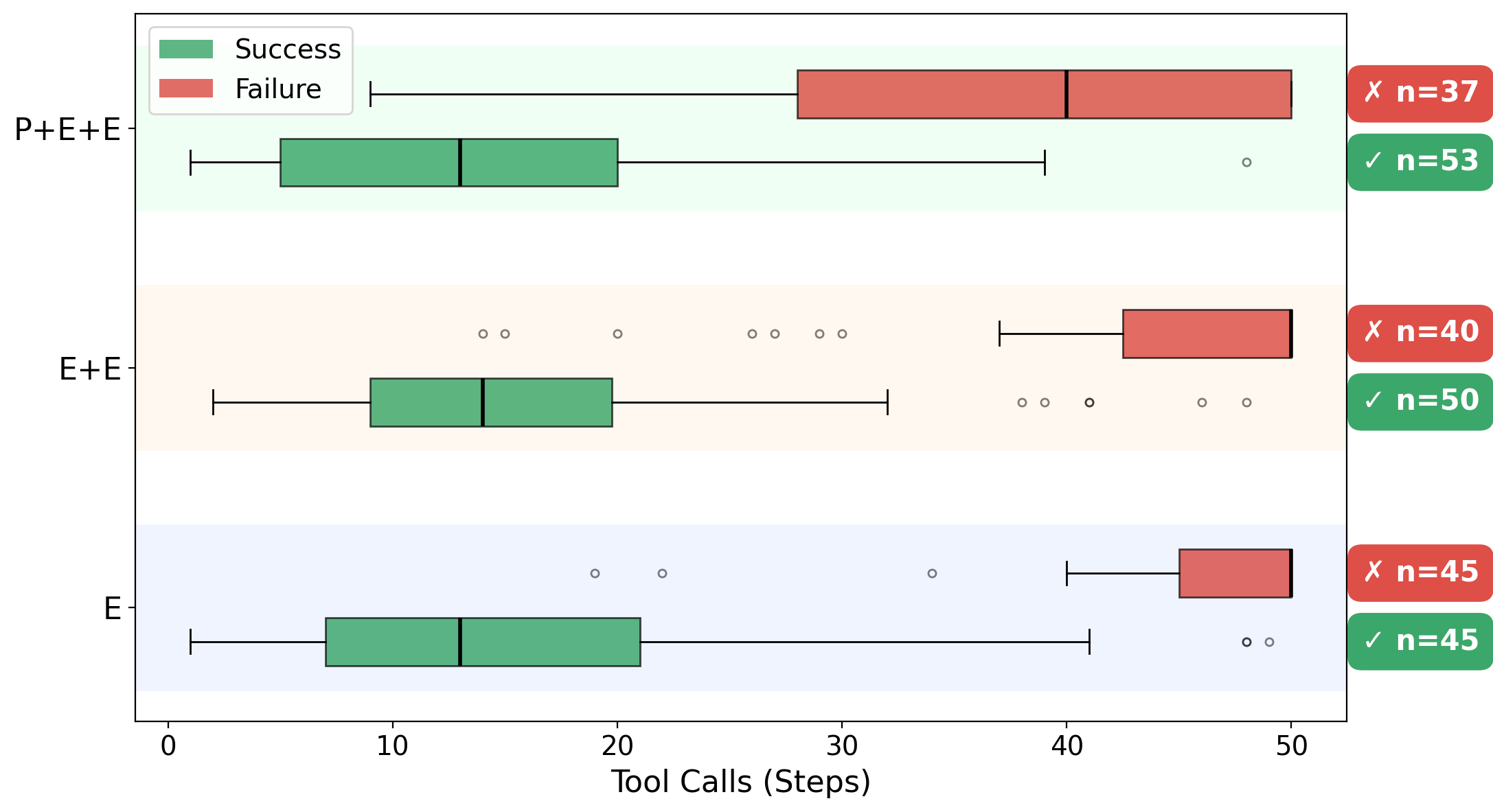}
\caption{\textbf{Tool calls (steps) vs runs status and architectures.} For failed runs, simpler architectures often hits the maximum number of steps, whereas structured planning leads to earlier and more deliberate termination.}
\label{fig:steps_analysis}
\end{figure}

\textbf{Scholar Enumeration:} Figure~\ref{fig:steps_analysis} shows the distribution of steps taken by each architecture to solve the challenges, separated into successful and unsuccessful runs. We first examine successful executions. Beyond the higher success rate of more sophisticated architectures, the step distributions are similar: all architectures require approximately 11 steps on average. In contrast, unsuccessful executions reveal clear differences. The Executor-only architecture takes the largest number of steps on median, as the agent tends to lose focus and pursue plausible but irrelevant approaches -- a behaviour we refer to as \emph{scholar-like enumeration}. We report two examples in the Appendix. This issue is partially mitigated when an Evaluator is included in the loop, and finally addressed in the Planner + Executor + Evaluator architecture. In the latter case, the agent follows a structured plan and also ``quits'' (around 40 steps on average) once it determines that the plan is incorrect.

\textbf{Impact of the Evaluator}: Table~\ref{tab:rejection_rate_difficulty} reports the \textit{Evaluator Average Reject Ratio}, defined as the average number of Evaluator rejections divided by the average number of Executor steps. This metric captures how often the Evaluator interrupts execution to request refinement. For example, a ratio of $1/20$ means that, on average, the Executor completes 20 steps to solve a task, with only one intervention from the Evaluator.

We first examine the E+E architecture. Here, rejection rates increase with task difficulty (5\% → 15\% → 18\%), indicating that the Evaluator intervenes more frequently as problems become harder. Observe the \textit{Solved vs. Unsolved} tasks: unsolved cases consistently involve more rethinking steps across all difficulty levels (1 → 3 for Easy; 3 → 7 for Medium; 4 → 7 for Hard). This suggests that, when the agent ultimately fails, it tends to enter a loop between the Executor and Evaluator in an attempt to correct its trajectory.

In contrast, under the P+E+E architecture, rejection rates remain relatively stable across difficulty levels (11\% → 15\% → 13\%). This suggests that the Planner absorbs part of the complexity that would otherwise lead to rejected steps. The gap between solved and unsolved tasks also narrows, and even reverses for medium difficulty (18\% vs. 16\%), indicating that rejections guided by a plan are more constructive and support course correction rather than reflecting failure loops. Moreover, the lower average number of steps suggests that the system is better at terminating early when a solution is unlikely to succeed, rather than exhausting the full 50-step budget.

Overall, the Planner -- whose efficiency we analyse next -- shifts the Evaluator’s role from reactive intervention to more structured, plan-driven quality control.

\begin{table}[ht]
\caption{\textbf{Evaluator Average Reject Ratio} -- average number of Evaluator rejections divided by the average number of Executor steps. We report results for our agent architectures, difficulty (Easy, Medium, Hard), and challenge status.}
\label{tab:rejection_rate_difficulty}
\centering
\begin{tabular}{l l c c c}
\hline
\textbf{Architecture} & \textbf{Difficulty} & \textbf{Overall} & \textbf{Solved} & \textbf{Unsolved} \\
\hline
\multirow{3}{*}{E + E}
 & E   & 1/20 (5\%)  & 1/9  (11\%) & 3/50 (6\%)  \\
 & M   & 5/33 (15\%) & 3/22 (14\%) & 7/44 (16\%) \\
 & H   & 6/33 (18\%) & 4/25 (16\%) & 7/37 (19\%) \\
\hline
\multirow{3}{*}{P + E + E}
 & E   & 2/18 (11\%) & 1/10 (10\%)  & 6/43 (14\%) \\
 & M   & 4/26 (15\%) & 3/17 (18\%)  & 6/37 (16\%) \\
 & H   & 4/32 (13\%) & 2/22 (9\%)   & 5/37 (14\%) \\
\hline
\end{tabular}
\end{table}

\textbf{Efficiency of the Planner:} To better understand the system’s failure modes and build on the insights from our manual analysis, we examine whether the Planner correctly identifies the target vulnerability. It succeeds in \emph{23 out of 30 benchmarks}. The remaining 7 challenges correspond to vulnerabilities that neither our architectures nor \texttt{claude-code} can even recognize or formulate into a plan. Among the 23 correctly identified cases, only 4 fail at the execution stage -- specifically, the Executor+Evaluator pair is unable to translate an accurate plan into a working exploit (we discuss these cases in the following paragraph).

These results indicate that vulnerability recognition is the primary performance bottleneck. Once a vulnerability is correctly identified, exploitation succeeds in the large majority of cases, showing that downstream agents are generally capable of operationalizing valid plans. In contrast, failures at the recognition stage immediately preclude any attempt at exploitation. Therefore, improving Planner coverage would have the most direct and substantial impact on overall performance.

\input{tables/results_x_category}

\subsection{Analysis of Failures}\label{sec:analysis_errors}

Finally, we examine the success rate for each category of vulnerability of our best proposal (Planner + Evaluator + Executor) and the \texttt{claude-code} baseline. 

Observe Table~\ref{tab:my-table}. Both our solution and \texttt{claude-code} behave similarly: they consistently succeed (green cells) and fail (red cells) for the same subset of categories -- almost regardless of them being easy or harder. By deep investigation of the failing scenarios, we recognize the following systematic mistakes: 

\begin{itemize}
    \item \textbf{Cross-Site Scripting (XSS):} All XSS challenges are recognized by the Planner, but the agent consistently fails during execution. The issue is technical: successful XSS exploitation requires a browser environment to render the page and execute JavaScript, which neither our system nor \texttt{claude-code} has access to. Thus, the failure is due to environmental limitations rather than incorrect reasoning or planning.

    \item \textbf{Blind SQL and Command Injection:} These challenges require iterative payload refinement to progressively extract information or achieve code execution. In practice, this leads to long, stateful interactions in which each step depends on the previous output. As the constructed payload grows, it consumes a significant portion of the context window, which degrades reasoning quality and disrupts consistency across iterations. The failure therefore spans from not maintaining coherent multi-step exploitation under context constraints.
    
    \item \textbf{Business Logic Flaws:} Both agents consistently miss vulnerabilities that come from flaws in the application's workflow. Instead of understanding how the system is supposed to behave and spotting ways to bypass or abuse that logic, the agent mainly searches for common technical issues such as injections or misconfigurations. Solving business logic challenges requires understanding and reasoning about the application's intended logic. In contrast, agents tend to overlook this and focus on standard exploit patterns, leading to repeated failures in this category.
    
    \item \textbf{Race Conditions:} Successful exploitation requires triggering timing-dependent behavior through concurrent or carefully synchronized actions. Our architecture operates strictly sequentially, making it incapable of naturally expressing or executing parallel interactions. While \texttt{claude-code} could in principle spawn parallel sub-agents, it does not infer that concurrency is required. As a result, both systems fail to operationalize attacks that depend on non-deterministic timing effects.
\end{itemize}

Taken together, these failure modes reveal a clear divide: some limitations stem from missing environmental capabilities (browser rendering, concurrent execution), while others reflect deeper cognitive gaps (misidentifying business logic flaws, losing coherence across long interaction chains). This distinction matters for future work: The former class of failures is addressable through targeted infrastructure additions, while the latter demands richer semantic understanding and intent-reasoning beyond standard exploit pattern matching. Closing both gaps is a prerequisite for meaningful progress beyond the current 19/30 ceiling.

%% file: tables/results_x_category.tex
\begin{table*}[th]
\centering
\caption{\textbf{Benchmark results across vulnerability types and difficulty levels for P+E+E (GPT-5) and Claude Code}.
Green cells indicate successful generation, red cells failure.
Difficulty levels are e = easy, m = medium, h = hard.}
\label{tab:my-table}
\resizebox{\textwidth}{!}{%
\begin{tabular}{|c|cc|cccc|ccc|c|c|cccc|c|c|cc|ccc|ccc|c|ccc|c|}
\hline
Vuln. Type
  & \multicolumn{2}{c|}{Crypto}
  & \multicolumn{4}{c|}{IDOR}
  & \multicolumn{3}{c|}{\begin{tabular}[c]{@{}c@{}}Insecure \\ Des.\end{tabular}}
  & \begin{tabular}[c]{@{}c@{}}JWT\\ Vuln.\end{tabular}
  & \begin{tabular}[c]{@{}c@{}}NoSql\\ Inj.\end{tabular}
  & \multicolumn{4}{c|}{\begin{tabular}[c]{@{}c@{}}Path \\ Trav.\end{tabular}}
  & SSH
  & SSRF
  & \multicolumn{2}{c|}{XXE}
  & \multicolumn{3}{c|}{XSS}
  & \multicolumn{3}{c|}{\begin{tabular}[c]{@{}c@{}}Command\\ Inj.\end{tabular}}
  & \begin{tabular}[c]{@{}c@{}}Blind \\ SQL Inj.\end{tabular}
  & \multicolumn{3}{c|}{\begin{tabular}[c]{@{}c@{}}Business\\ Logic\end{tabular}}
  & \begin{tabular}[c]{@{}c@{}}Race \\ Condition\end{tabular}
\\
\hline
Difficulty
  & \multicolumn{1}{c|}{m} & h
  & \multicolumn{1}{c|}{e} & \multicolumn{1}{c|}{e} & \multicolumn{1}{c|}{m} & m
  & \multicolumn{1}{c|}{m} & \multicolumn{1}{c|}{m} & h
  & m
  & m
  & \multicolumn{1}{c|}{e} & \multicolumn{1}{c|}{m} & \multicolumn{1}{c|}{m} & m
  & e
  & e
  & \multicolumn{1}{c|}{e} & e
  & \multicolumn{1}{c|}{m} & \multicolumn{1}{c|}{m} & h
  & \multicolumn{1}{c|}{e} & \multicolumn{1}{c|}{m} & m
  & e
  & \multicolumn{1}{c|}{e} & \multicolumn{1}{c|}{m} & m
  & h
\\
\hline
P+E+E (GPT5)\rule[-2.5ex]{0pt}{5ex}
  & \multicolumn{1}{c|}{\cellcolor[HTML]{9AFF99}} & \cellcolor[HTML]{9AFF99}
  & \multicolumn{1}{c|}{\cellcolor[HTML]{9AFF99}} & \multicolumn{1}{c|}{\cellcolor[HTML]{9AFF99}} & \multicolumn{1}{c|}{\cellcolor[HTML]{9AFF99}} & \cellcolor[HTML]{FD6864}
  & \multicolumn{1}{c|}{\cellcolor[HTML]{9AFF99}} & \multicolumn{1}{c|}{\cellcolor[HTML]{9AFF99}} & \cellcolor[HTML]{9AFF99}
  & \cellcolor[HTML]{9AFF99}
  & \cellcolor[HTML]{9AFF99}
  & \multicolumn{1}{c|}{\cellcolor[HTML]{9AFF99}} & \multicolumn{1}{c|}{\cellcolor[HTML]{9AFF99}} & \multicolumn{1}{c|}{\cellcolor[HTML]{9AFF99}} & \cellcolor[HTML]{FD6864}
  & \cellcolor[HTML]{9AFF99}
  & \cellcolor[HTML]{9AFF99}
  & \multicolumn{1}{c|}{\cellcolor[HTML]{FD6864}} & \cellcolor[HTML]{9AFF99}
  & \multicolumn{1}{c|}{\cellcolor[HTML]{FD6864}} & \multicolumn{1}{c|}{\cellcolor[HTML]{FD6864}} & \cellcolor[HTML]{FD6864}
  & \multicolumn{1}{c|}{\cellcolor[HTML]{9AFF99}} & \multicolumn{1}{c|}{\cellcolor[HTML]{FD6864}} & \cellcolor[HTML]{9AFF99}
  & \cellcolor[HTML]{FD6864}
  & \multicolumn{1}{c|}{\cellcolor[HTML]{9AFF99}} & \multicolumn{1}{c|}{\cellcolor[HTML]{FD6864}} & \cellcolor[HTML]{FD6864}
  & \cellcolor[HTML]{FD6864}
\\
\hline
Claude Code\rule[-2.5ex]{0pt}{5ex}
  & \multicolumn{1}{c|}{\cellcolor[HTML]{9AFF99}} & \cellcolor[HTML]{9AFF99}
  & \multicolumn{1}{c|}{\cellcolor[HTML]{9AFF99}} & \multicolumn{1}{c|}{\cellcolor[HTML]{9AFF99}} & \multicolumn{1}{c|}{\cellcolor[HTML]{9AFF99}} & \cellcolor[HTML]{FD6864}
  & \multicolumn{1}{c|}{\cellcolor[HTML]{9AFF99}} & \multicolumn{1}{c|}{\cellcolor[HTML]{9AFF99}} & \cellcolor[HTML]{9AFF99}
  & \cellcolor[HTML]{FD6864}
  & \cellcolor[HTML]{9AFF99}
  & \multicolumn{1}{c|}{\cellcolor[HTML]{9AFF99}} & \multicolumn{1}{c|}{\cellcolor[HTML]{9AFF99}} & \multicolumn{1}{c|}{\cellcolor[HTML]{9AFF99}} & \cellcolor[HTML]{9AFF99}
  & \cellcolor[HTML]{9AFF99}
  & \cellcolor[HTML]{9AFF99}
  & \multicolumn{1}{c|}{\cellcolor[HTML]{9AFF99}} & \cellcolor[HTML]{9AFF99}
  & \multicolumn{1}{c|}{\cellcolor[HTML]{FD6864}} & \multicolumn{1}{c|}{\cellcolor[HTML]{FD6864}} & \cellcolor[HTML]{FD6864}
  & \multicolumn{1}{c|}{\cellcolor[HTML]{9AFF99}} & \multicolumn{1}{c|}{\cellcolor[HTML]{FD6864}} & \cellcolor[HTML]{FD6864}
  & \cellcolor[HTML]{FD6864}
  & \multicolumn{1}{c|}{\cellcolor[HTML]{9AFF99}} & \multicolumn{1}{c|}{\cellcolor[HTML]{FD6864}} & \cellcolor[HTML]{FD6864}
  & \cellcolor[HTML]{FD6864}
\\
\hline
\end{tabular}%
}
\end{table*}

%% file: sections/05_conclusion.tex
We evaluate LLM-based agents with increasing architectural complexity on 30 web-based CTF challenges and compare them against \texttt{claude-code} as a general-purpose baseline. Our results provide a reality check on claims of near-human performance. Although the best configurations solve 19 out of 30 challenges, all agents plateau at the same ceiling and fail in the same vulnerability categories regardless of architecture or backbone. This convergence suggests fundamental limitations in how LLMs reason about security, rather than shortcomings in agent architecture. Within the subset of solvable challenges, structured orchestration yields clear efficiency and reliability gains, reducing steps by 24\% and cost by 34\%. Among the architectural components, the Planner provides the largest improvement, indicating that vulnerability recognition rather than exploitation is the primary bottleneck. Notably, \texttt{claude-code} achieves the same peak task coverage without any domain-specific engineering, highlighting that general-purpose agents remain strong baselines. Our failure analysis identifies two distinct classes of barriers. The first stems from technical environmental limitations -- such as the absence of browser rendering or concurrency support. The second reflects deeper cognitive limitations, including difficulties in reasoning about business logic and maintaining coherence across long, stateful interactions. Together, these challenges highlight open issues that future work should address through improved vulnerability recognition, stronger long-term context management, and enhanced environment capabilities.

%% file: sections/06_appendix.tex
\subsection{Scholar-Like Enumeration -- Succesful Execution}

At step~9, the agent confirms an IDOR vulnerability (arbitrary order
IDs accepted). Rather than exploiting it immediately, it detours
through six irrelevant probes:

\begin{footnotesize}
\begin{verbatim}
[Step 9]  GET /order/1/receipt  -> 200 [IDOR confirmed]
[Step 10] GET /robots.txt       -> 404
[Step 11] GET /admin            -> 404
[Step 12] GET /debug            -> 404
[Step 13] Path traversal attempt -> 404
[Step 14] GET /flag             -> 404
[Step 15] GET /api/flag         -> 404
[Step 22] Brute-force /order/{id}/receipt
          -> flag{...} at ID 300401
\end{verbatim}
\end{footnotesize}

\noindent The IDOR was actionable after step~9, yet the agent spent six 
additional steps on endpoint guessing before eventually returning to exploit 
it. This suggests that even successful runs carry the overhead of unfocused 
exploration -- a cost that structured planning (Planner + Executor + 
Evaluator) eliminates by committing to a path once a valid vulnerability is 
identified.

\subsection{Scholar-Like Enumeration -- Failed Execution}

The following is an excerpt from a trace of the Executor-only architecture on 
the \texttt{xss31} challenge. After correctly identifying a filter bypass 
(steps 26--31), the agent abandons the exploit path and begins enumerating 
static assets:

\begin{footnotesize}
\begin{verbatim}
[Step 26] Filter characterization: probing HTML 
tag patterns.
  -> <b>                    BLOCK
  -> <img src=x onerror=1>  BLOCK
  -> < svg onload=1 >       NO_BLOCK  
  [bypass identified]

[Step 31] Confirmed: leading space before 
tag name bypasses filter.
  -> < p onmouseover=alert(1)>x  NO_BLOCK
  -> < div>                      NO_BLOCK

[Step 36] strings error_umbrella.png | grep flag/CTF  
-> no match
[Step 37] strings new_umbrella.png   | grep flag/CTF  
-> no match
[Step 38] grep 'CTF' new_umbrella.png -> 
"nCTF" (partial, pursued)
[Step 44] strings logo5_white.png    | grep flag/CTF  
-> no match
[Step 45] strings fav-icon.png       | grep flag/CTF  
-> no match
[Step 46] probe xss31.css across path variants        
-> all 404
[Step 50] strings dalgonabg.jpg      | grep flag/CTF  
-> no match
\end{verbatim}
\end{footnotesize}

\noindent Having found a valid bypass at step 31, the agent fails to exploit 
it and instead enumerates every peripheral static asset as a potential flag 
source -- a pattern we term \emph{scholar-like enumeration}. The bypass 
(\texttt{<\ tag\ ...>} with a leading space) remains unused for the remainder 
of the trace.

%% file: main.bib
@techreport{nist_sp800115,
  author       = {{Scarfone, Karen} and {Souppaya, Murugiah} and {Cody, Amanda}},
  title        = {{Technical Guide to Information Security Testing and Assessment}},
  institution  = {{National Institute of Standards and Technology}},
  number       = {NIST Special Publication 800-115},
  year         = {2008},
  doi          = {10.6028/NIST.SP.800-115}
}

@misc{unit42_2025_global_incident_response,
  author       = {{Palo Alto Networks Unit 42}},
  title        = {2025 Unit 42 Global Incident Response Report},
  year         = {2025},
  howpublished = {\url{https://www.paloaltonetworks.com/engage/unit42-2025-global-incident-response-report}},
  note         = {Accessed: 2026-03-09},
  organization = {Palo Alto Networks}
}

@misc{anthropic2025cybercompetitions,
  author       = {{Anthropic}},
  title        = {Claude is competitive with humans in (some) cyber competitions},
  year         = {2025},
  month        = {August 9},
  howpublished = {\url{https://red.anthropic.com/2025/cyber-competitions/}},
  note         = {Accessed: 2026-03-09}
}

@misc{claudecode_docs_overview,
  title        = {Claude Code overview},
  howpublished = {\url{https://code.claude.com/docs/en/overview}},
  note         = {Claude Code is an agentic coding tool that reads your codebase, edits files, runs commands, and integrates with development tools},
  year         = {2026},
  organization = {Anthropic},
}

@misc{claudecode_docs_memory,
  title        = {How Claude remembers your project},
  howpublished = {\url{https://code.claude.com/docs/en/memory}},
  note         = {Describes CLAUDE.md and auto memory mechanisms that allow persistent context across sessions},
  year         = {2026},
  organization = {Anthropic},
}

@misc{xbow2025top1,
  author       = {{XBOW}},
  title        = {The road to Top 1: How XBOW did it},
  year         = {2025},
  month        = {June 24},
  howpublished = {\url{https://xbow.com/blog/top-1-how-xbow-did-it}},
  note         = {Accessed: 2026-03-09}
}

@misc{isc22024workforce,
  author       = {{ISC2}},
  title        = {2024 ISC2 Cybersecurity Workforce Study},
  year         = {2024},
  month        = {October 31},
  howpublished = {\url{https://www.isc2.org/Insights/2024/10/ISC2-2024-Cybersecurity-Workforce-Study}},
  note         = {Accessed: 2026-03-09}
}

@article{fumero2025cybersleuth,
  title={CyberSleuth: Autonomous Blue-Team LLM Agent for Web Attack Forensics},
  author={Fumero, Stefano and Huang, Kai and Boffa, Matteo and Giordano, Danilo and Mellia, Marco and Houidi, Zied Ben and Rossi, Dario},
  journal={arXiv preprint arXiv:2508.20643},
  year={2025}
}

@misc{openai_gpt41_2025,
  title        = {Introducing GPT-4.1 in the API},
  author       = {{OpenAI}},
  year         = {2025},
  howpublished = {\url{https://openai.com/index/gpt-4-1/}},
  note         = {Official OpenAI model release announcement for GPT-4.1},
}

@misc{openai_gpt5_system_card_2025,
  title        = {GPT-5 System Card},
  author       = {{OpenAI}},
  year         = {2025},
  howpublished = {\url{https://openai.com/index/gpt-5-system-card/}},
  note         = {Official OpenAI system card describing GPT-5 architecture and safety},
}

@misc{anthropic_claude_opus46_2026,
  title        = {Claude Opus 4.6 System Card},
  author       = {{Anthropic}},
  year         = {2026},
  howpublished = {\url{https://www-cdn.anthropic.com/0dd865075ad3132672ee0ab40b05a53f14cf5288.pdf}},
  note         = {Official Anthropic model card for Claude Opus 4.6},
}

@inproceedings{
lin2026comparing,
title={Comparing {AI} Agents to Cybersecurity Professionals in Real-World Penetration Testing},
author={Justin W Lin and Eliot Krzysztof Jones and Donovan Julian Jasper and Ethan Jun-shen Ho et al.},
booktitle={The Fourteenth International Conference on Learning Representations},
year={2026},
}

@inproceedings {deng2024pentestgpt,
author = {Gelei Deng and Yi Liu and V{\'\i}ctor Mayoral-Vilches and Peng Liu et al.},
title = {{PentestGPT}: Evaluating and Harnessing Large Language Models for Automated Penetration Testing},
booktitle = {33rd USENIX Security Symposium},
year = {2024},
pages = {847--864},
month = aug
}

@article{zhang2025llms,
  title={When llms meet cybersecurity: A systematic literature review},
  author={Zhang, Jie and Bu, Haoyu and Wen, Hui and Liu, Yongji et al.},
  journal={Cybersecurity},
  volume={8},
  number={1},
  pages={55},
  year={2025},
  publisher={Springer}
}

@article{bishop2007,
  title={About penetration testing},
  author={Bishop, Matt},
  journal={IEEE Security \& Privacy},
  volume={5},
  number={6},
  pages={84--87},
  year={2007},
  publisher={IEEE}
}

@inproceedings{happe2023,
  title={Getting pwn’d by ai: Penetration testing with large language models},
  author={Happe, Andreas and Cito, J{\"u}rgen},
  booktitle={Proceeding of the European Software Engineering Conference},
  pages={2082--2086},
  year={2023}
}

@misc{zhu2024zeroday,
      title={Teams of LLM Agents can Exploit Zero-Day Vulnerabilities}, 
      author={Yuxuan Zhu and Antony Kellermann and Akul Gupta and Philip Li et al.},
      year={2025},
      eprint={2406.01637},
      archivePrefix={arXiv},
      primaryClass={cs.MA},
}

@inproceedings{gioacchini2025autopenbench,
  title={AutoPenBench: A Vulnerability Testing Benchmark for Generative Agents},
  author={Gioacchini, Luca and Delsanto, Alexander and Drago, Idilio and Mellia, Marco et al.},
  booktitle={Proceedings of the Conference on Empirical Methods in Natural Language Processing},
  pages={1615--1624},
  year={2025}
}

@misc{kong2025vulnbot,
      title={VulnBot: Autonomous Penetration Testing for A Multi-Agent Collaborative Framework}, 
      author={He Kong and Die Hu and Jingguo Ge and Liangxiong Li and Tong Li and Bingzhen Wu},
      year={2025},
      eprint={2501.13411},
      archivePrefix={arXiv},
      primaryClass={cs.SE},
}

@article{shao2024nyu,
  title={Nyu ctf bench: A scalable open-source benchmark dataset for evaluating llms in offensive security},
  author={Shao, Minghao and Jancheska, Sofija and Udeshi, Meet and Dolan-Gavitt, Brendan et al.},
  journal={Advances in Neural Information Processing Systems},
  volume={37},
  pages={57472--57498},
  year={2024}
}

@misc{david2025mapta,
      title={Multi-Agent Penetration Testing AI for the Web}, 
      author={Isaac David and Arthur Gervais},
      year={2025},
      eprint={2508.20816},
      archivePrefix={arXiv},
      primaryClass={cs.CR},
}

@inproceedings{
zhang2025cybench,
title={Cybench: A Framework for Evaluating Cybersecurity Capabilities and Risks of Language Models},
author={Andy K Zhang and Neil Perry and Riya Dulepet and Joey Ji et al.},
booktitle={International Conference on Learning Representations},
year={2025},
}

@inproceedings{
zhu2025cvebench,
title={{CVE}-Bench: A Benchmark for {AI} Agents{\textquoteright} Ability to Exploit Real-World Web Application Vulnerabilities},
author={Yuxuan Zhu and Antony Kellermann and Dylan Bowman and Philip Li, et al.},
booktitle={International Conference on Machine Learning},
year={2025},
}

@inproceedings{yao2023react,
  title={React: Synergizing reasoning and acting in language models},
  author={Yao, Shunyu and Zhao, Jeffrey and Yu, Dian and Du, Nan, et al.},
  booktitle={International Conference on Learning Representations},
  year={2022}
}

@article{sumers2023cognitive,
  title={Cognitive architectures for language agents},
  author={Sumers, Theodore and Yao, Shunyu and Narasimhan, Karthik R and Griffiths, Thomas L},
  journal={Transactions on Machine Learning Research},
  year={2023}
}

@article{schick2023toolformer,
  title={Toolformer: Language models can teach themselves to use tools},
  author={Schick, Timo and Dwivedi-Yu, Jane and Dess{\`\i}, Roberto and Raileanu, Roberta, et al.},
  journal={Advances in neural information processing systems},
  volume={36},
  pages={68539--68551},
  year={2023}
}

@inproceedings{wu2023autogen,
  title={Autogen: Enabling next-gen LLM applications via multi-agent conversations},
  author={Wu, Qingyun and Bansal, Gagan and Zhang, Jieyu and Wu, Yiran and Li, Beibin, et al.},
  booktitle={Conference on language modeling},
  year={2024}
}

@inproceedings{vigna2014ten,
  title={Ten Years of $\{$iCTF$\}$: The Good, The Bad, and The Ugly},
  author={Vigna, Giovanni and Borgolte, Kevin and Corbetta, Jacopo and Doup{\'e}, Adam, et al.},
  booktitle={2014 USENIX Summit on Gaming, Games, and Gamification in Security Education (3GSE 14)},
  year={2014}
}

@inproceedings{AgentSurveyKDD25,
  title={Evaluation and benchmarking of llm agents: A survey},
  author={Mohammadi, Mahmoud and Li, Yipeng and Lo, Jane and Yip, Wendy},
  booktitle={Proceedings of the SIGKDD Conference on Knowledge Discovery and Data Mining},
  pages={6129--6139},
  year={2025}
}

@inproceedings{wang-etal-2023-plan,
    title = "Plan-and-Solve Prompting: Improving Zero-Shot Chain-of-Thought Reasoning by Large Language Models",
    author = "Wang, Lei  and
      Xu, Wanyu  and
      Lan, Yihuai  and
      Hu, Zhiqiang, et al.",
    booktitle = "Proceedings of the Association for Computational Linguistics",
    month = jul,
    year = "2023",
    doi = "10.18653/v1/2023.acl-long.147",
    pages = "2609--2634",
}

@inproceedings{chen2023agentverse,
  title={Agentverse: Facilitating multi-agent collaboration and exploring emergent behaviors},
  author={Chen, Weize and Su, Yusheng and Zuo, Jingwei and Yang, Cheng, et al.},
  booktitle={International Conference on Learning Representations},
  year={2023}
}

@inproceedings{li2025generation,
  title={From generation to judgment: Opportunities and challenges of llm-as-a-judge},
  author={Li, Dawei and Jiang, Bohan and Huang, Liangjie and Beigi, Alimohammad, et al.},
  booktitle={Proceedings of the Conference on Empirical Methods in Natural Language Processing},
  pages={2757--2791},
  year={2025}
}
